\documentclass[a4paper,11pt]{article}
\usepackage{arxiv}
\usepackage{amsmath,amsfonts,amssymb}
\usepackage{graphicx}
\usepackage[colorlinks=true, allcolors=blue]{hyperref}
\usepackage{threeparttable} % to use table notes
\usepackage{macro_ref}
\usepackage{geometry}
\usepackage{natbib}

\title{Generic data reduction for nulling interferometry package: the grip of a single data reduction package on all the nulling interferometers}
\renewcommand{\undertitle}{}
\renewcommand{\headeright}{}
\renewcommand{\shorttitle}{GRIP}
\date{}

\begin{document} 
\maketitle

\begin{center}
\vspace{-5em}
Marc-Antoine Martinod$^a$\footnote{marc-antoine.martinod@kuleuven.be},
Denis Defrère$^a$,
Romain Laugier$^a$,
Steve Ertel$^{d, e}$,
Olivier Absil$^c$,
Barnaby Norris$^b$,
Germain Garreau$^a$,
Bertrand Mennesson$^f$.
~\newline
~\newline
$^a$Institute of Astronomy, KU Leuven, Celestijnenlaan 200D, 3001 Leuven, Belgium;
$^b$Sydney Institute for Astronomy, School of Physics, Physics Road, University of Sydney, NSW 2006, Australia;
$^c$Space sciences, Technologies \& Astrophysics Research (STAR) Institute, University of Li\`ege, Li\`ege, Belgium;
$^d$Department of Astronomy and Steward Observatory, 933 North Cherry Ave, Tucson, AZ 89 85721, USA;
$^e$Large Binocular Telescope Observatory, 933 North Cherry Ave, Tucson, AZ 85721, USA;
$^f$Jet Propulsion Laboratory, California Institute of Technology (United States).

\end{center}
~\newline

\begin{abstract}
Nulling interferometry is a powerful observing technique to reach exoplanets and circumstellar dust at separations too small for direct imaging with single-dish telescopes and too large for indirect methods. 
With near-future instrumentation, it bears the potential to detect young, hot planets near the snow lines of their host stars.  
A future space mission could detect and characterize a large number of rocky, habitable-zone planets around nearby stars at thermal-infrared wavelengths.
The null self-calibration is a method aiming at modelling the statistical distribution of the nulled signal. 
It has proven to be more sensitive and accurate than average-based data reduction methods in nulling interferometry. 
This statistical approach opens the possibility of designing a GPU-based Python package to reduce the data from any of these instruments, by simply providing the data and a simulator of the instrument. 
\texttt{GRIP} is a toolbox to reduce nulling and interferometric data based on the statistical self-calibration method. 
In this article, we present the main features of GRIP as well as applications on real data.
\end{abstract}

% Include a list of keywords after the abstract 

\section{Introduction}
Nulling interferometry is a powerful technique to perform high contrast imaging in the close neighbourhood of the stars \citep{bracewell1978}.
Pioneered on the Multiple-Mirror Telescope (MMT) in 1998 \citep{Hinz1998}, it has been used over the past decades to detect and characterize habitable-zone dust around nearby stars.
It is the equivalent of coronagraphy in interferometry: beams from two apertures are coherently combined to create a destructive interference for on-axis sources.
The offset in the optical path lengths induced by any off-axis sources (e.g.\ an exoplanet) is transmitted by the instrument, allowing a direct detection and characterisation of the object.
The nulled depth is linked to the observed target's spatial coherence (also known as the visibility) with $N = \frac{1-|\tilde{V}|}{1+|\tilde{V}|}$.
Performing nulling requires extremely stable interferences to efficiently discriminate the photons coming from the planets from the ones coming from the starlight.
The fainter the off-axis source, the better the stabilisation must be, to efficiently remove the starlight from the measured signal.
Such observation is challenged by atmospheric turbulence.

Legacy data reduction techniques relied on measuring the average value of the null depth on the science and calibrator targets then performed a subtraction between the two to obtain the astrophysical null depth, solely governed by the object's geometry.
The current method is the null self-calibration \citep{hanot2011}: it models the statistical fluctuations to provide a self-calibrated astrophysical null depth.
In other words, this method directly measures the null depth without the biases coming from the instrumental response.
Therefore, the measurement is more precise and the observation of a reference star to measure this response is ideally not required.

The null self-calibration is a simulation-based inference method.
Monte-Carlo simulations with a simulator of the instrument are carried out to reproduce the measured distributions.
This method generates sequences of quantities based on measured or assumed distributions of the sources of noise and the astrophysical null depth, and propagates them throughout the simulator to obtain a simulated sequence of nulled flux.
The generated data are sorted into a histogram, which is then fit to the histogram of the real data to find the best-fit parameters.
This process enables the self-calibration, providing the simulator can correctly model the dominant effects of the instrument response. 
In some cases, calibration with reference targets is required to remove unknown sources of null errors.
One of the parameters is the astrophysical null depth, i.e. the observable which only depends on the surface brightness and the geometry of the target, free from the bias of the instrument response.
The use of this technique on the Palomar Fiber Nuller yielded an improvement on the precision by a factor of 10 \citep{hanot2011, Mennesson2011, serabyn2019}.
This method is now used on active nullers such as the LBTI \citep{defrere2016, Mennesson2016} and GLINT \citep{norris2020, martinod2021}.
It has enabled several scientific achievements on spectroscopic binaries \citep{serabyn2019}, exozodiacal dust \citep{ertel2020aj} and stellar diameters \citep{martinod2021}.

By relying on a Monte-Carlo approach and a clearly defined instrumental function, this technique can be used on any nuller.
It shifts the way of doing data reduction in interferometry from a tailor-made pipeline for each instrument to a generic one providing self-calibrated measurements.
The use of a generic pipeline is routinely used in direct imaging, e.g., with the \texttt{Vortex Imaging Processing} package \citep{gonzalez2017}.
This paper introduces the \emph{Generic data reduction for nulling interferometry package} (GRIP) which aims to provide a universal pipeline to all existing and future nuller.

\section{Goal and concept of GRIP}
The goal of \texttt{GRIP} is to provide self-calibrated null depth measurements by fitting a model of the instrumental perturbations to histograms of data.
The model is generated using a simulator of the instrument, which can be provided by the user or is built into the package for the main operating nullers (i.e. LBTI and GLINT).
\texttt{GRIP}'s model fitting strategies recover up to 3 parameters: the astrophysical null depth (or any other similar observable), the location and scale parameters of a normal distribution describing a particular noise in the simulator (e.g. phase fluctuations).
\texttt{GRIP} offers several optimisation strategies and cost functions to perform the fit.

\begin{figure}[hbtp]
    \centering
    \includegraphics[width=0.6\textwidth]{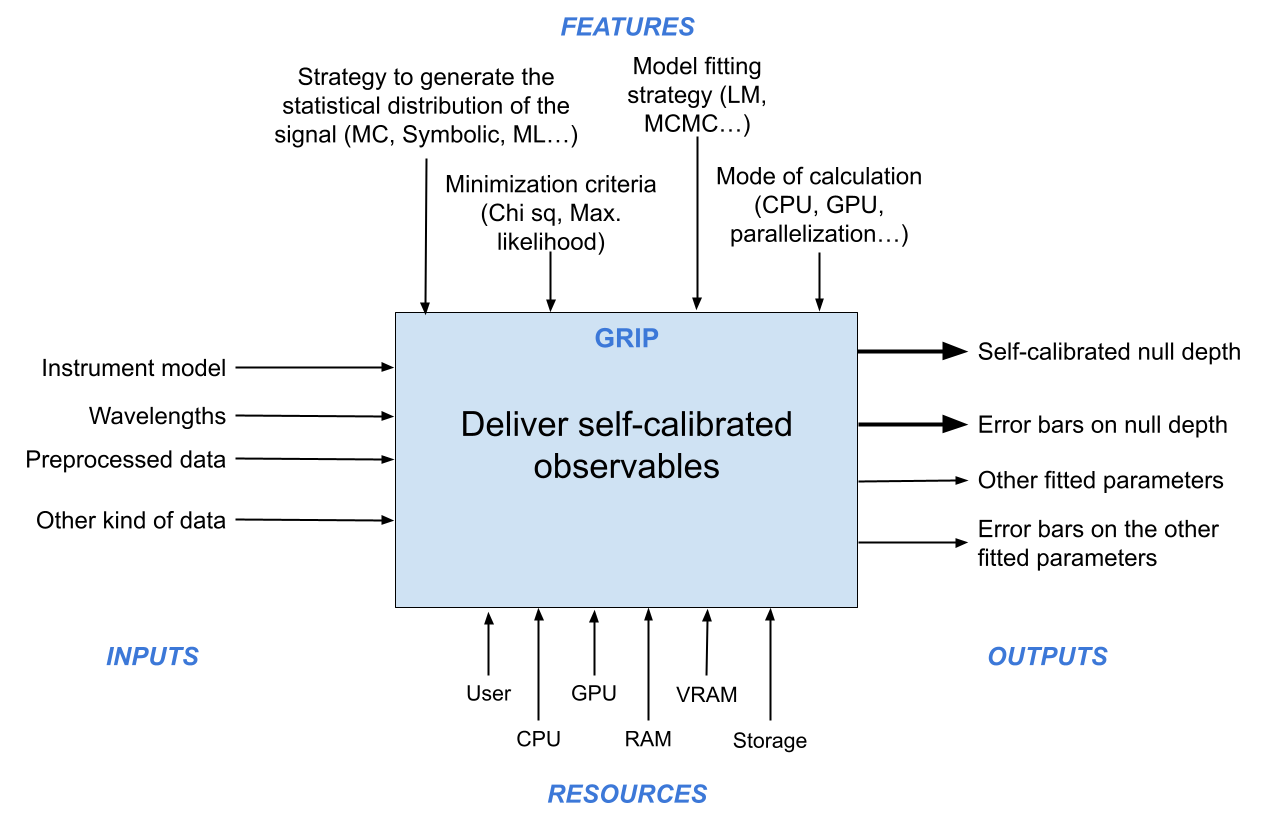}
    \caption{Structured analysis and design technique of GRIP. The left part lists the elements provided by the user and the right parts are the outputs of \texttt{GRIP}. The upper side lists the features of GRIP to process the data and the lower part lists the required hardware resources.}
    \label{fig:grip_sadt}
\end{figure}

Figure~\ref{fig:grip_sadt} details how this goal is achieved.
In the inputs, the user has to provide a simulator of the instrument (also called \textit{simulator}), any spectral information, preprocessed data to extract the astrophysical null depth and ancillary data to be used by the simulator.
Preprocessed data, consisting of sequences such as the flux of outputs or observables, are organized into histograms instead of time series.
The features are going to process the inputs to deliver the outputs.
They include optimisation strategies and cost functions.
The use of the package's features requires resources such as user supervision and computer components.
The outputs of \texttt{GRIP} are the self-calibrated null depth and the other fitted parameters in the form of their values and their uncertainties.

\texttt{GRIP} is a package designed to build a data reduction pipeline; it does not come as ready-to-use software.
However, it offers a supported and standardised way to reduce in the same way data from any type of nuller, providing the appropriate simulator.

\section{Package overview}
The code is being developed in \texttt{Python}, under the MIT license, and it is published on \texttt{Github}\footnote{\url{https://github.com/mamartinod/grip}}.
Its dependencies are widely-used Python packages such as \texttt{NumPy} and \texttt{scipy}.
\texttt{GRIP} is a function-based package although a class approach may not be excluded in the future.
The documentation is made of the internal documentation of each function, describing the aim, inputs and outputs of the function, and tutorials, explaining the use of the features of the package.
They are found on the project's Github repository.
Particularly, one tutorial is provided to help the design of an instrument simulator to use with \texttt{GRIP}.
\texttt{GRIP} runs on a GPU with the use of \texttt{Cupy} \citep{cupy_learningsys2017} or on CPU if this package is not installed.

\texttt{GRIP}'s features are used to build a data reduction pipeline.
These features are set in several modules (Table~\ref{tab:grip_modules}).
A Monte-Carlo process is used to generate a sequence of data, turned into a histogram, which is fit to the observed histogram using optimisers from well-known third-party Python packages.
The user defines the quantities to be generated during the Monte Carlo process and \texttt{GRIP} can efficiently reproduce any arbitrary distribution, allowing the use of the distributions of measured data.

\begin{table}[h]
    \centering
    \caption{Modules of GRIP}
    \begin{tabular}{ll}
    \hline
    \hline
        \texttt{fitting} & Fitting optimisers, estimators for fitting and a parameter space scanner \\
        \texttt{generic} & Miscellaneous functions \\
        \texttt{histogram\_tools} & Histograms makers and random values generators \\
        \texttt{instrument\_models} & Instrument models such as GLINT \citep{martinod2021} or LBTI \citep{defrere2016} \\
        \texttt{load\_files} & Load HDF5 and FITS files \\
        \texttt{plots} & Plot results of GRIP \\
        \texttt{preprocessing} & Features such as sorting and data binning \\
    \hline
    \end{tabular}
    \label{tab:grip_modules}
\end{table}

\texttt{GRIP} is made of several modules, providing all the tools to make histograms, perform self-calibration measurements, visualise the results and check the goodness of the fit.
The \texttt{fitting} and \texttt{histogram\_tools} are core modules of \texttt{GRIP}.
They contain all the utilities to perform null self-calibration (optimiser, cost function, random value generator, histogram maker) and can handle user-defined instrument simulators.
The current optimisers implemented in GRIP are: \texttt{scipy.optimize.least\_squares} with the dogbox method, \texttt{scipy.optimize.minimize} with the Powell method.
Several maximum likelihood estimators are provided: the logarithm of the binomial distribution, the sum of the squared residuals (like the one used in \texttt{scipy}'s least square algorithm), the Pearson's $\chi^2$ (coming from Pearson's chi-squared statistical test).
Some of the nullers are already implemented in  \texttt{instrument\_models} such as GLINT \citep{martinod2021} or LBTI \citep{Hinz2016, Ertel2020}.
Alongside optimisers and likelihood estimators, \texttt{fitting} includes a parameter space scanner.
This feature allows the exploration of the parameter space in a brute-force manner for the user to define the appropriate boundaries for the model fitting.

Functions in \texttt{histogram\_tools} create histograms from measured data or simulations, perform random values generation and deliver diagnostic data to help to assess the reliability of the fit.
The module \texttt{plots} includes functions to display the results of the fit, the parameter space as well as the diagnostic data.
The module \texttt{load\_files} includes functions to read HDF5 and FITS files; the user specifies the keywords to read, and the functions return a dictionary with the requested data.
These can be then preprocessed with the tools, such as frame binning, sorting, slicing, and extracting spectral or photometric information, present in the \texttt{preprocessing} module.

\section{GRIP in practice}
\begin{figure}[h]
    \centering
    \begin{tabular}{cc}
        \includegraphics[width=0.48\textwidth]{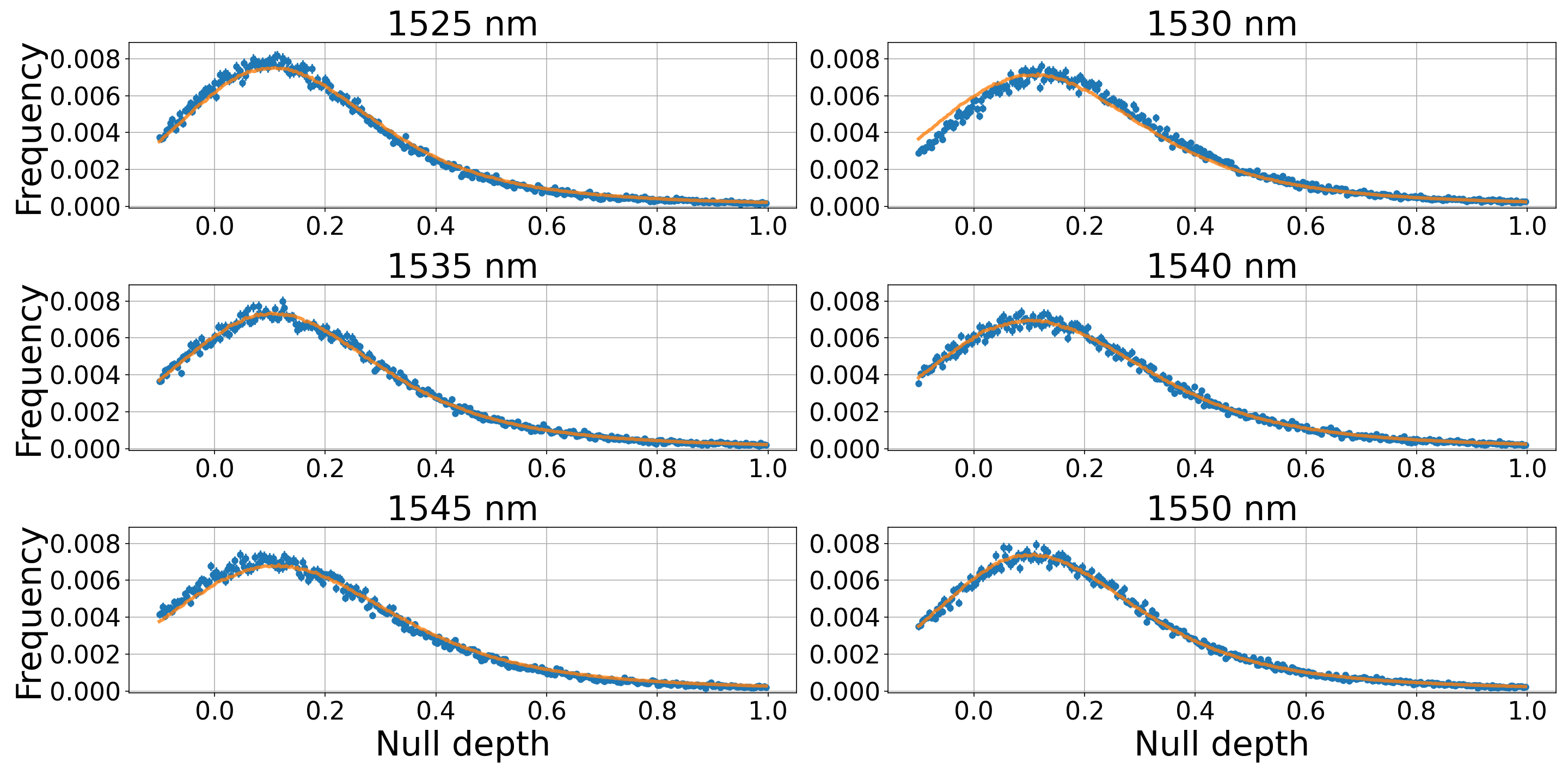} & 
        \includegraphics[width=0.48\textwidth]{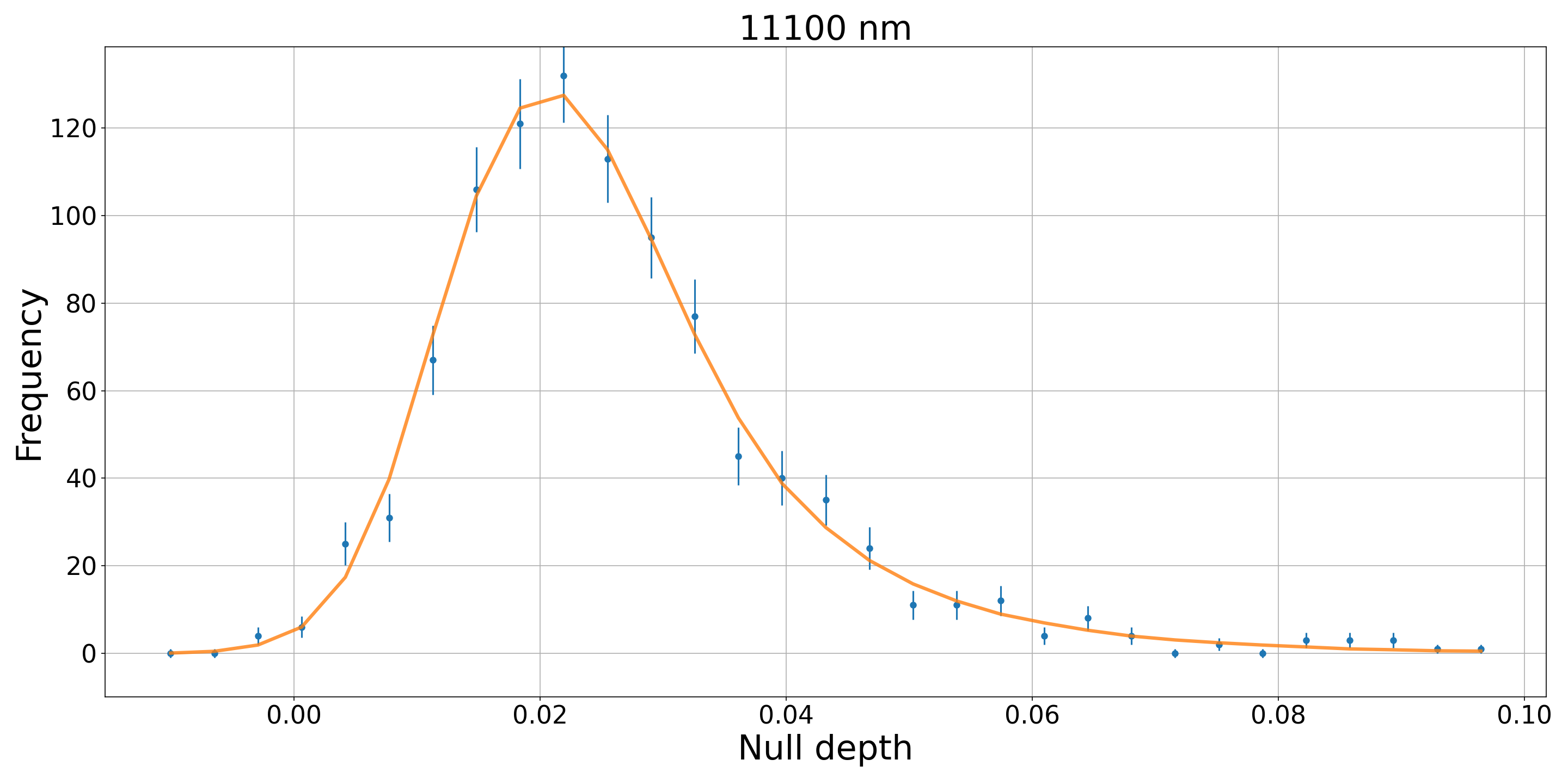} \\
    \end{tabular}
    \caption{Fitted histograms of GLINT (left) and LBTI (right) data with \texttt{GRIP}. Left: GLINT data are Arcturus, taken with a baseline of 5.55 meters. The histograms of the null depth is made for every spectral channel are fitted at once. The fitted values are: $N_a = 0.0705 \pm 0.000137$, $\mu_{OPD} = 302 \pm 7.34$~nm and $\sigma_{OPD} = 163 \pm 0.378$~nm. Right: LBTI data are one observing block of $\beta$~Leo, taken with a nominal baseline of 14.4 meters centre to centre, there is no spectral dispersion. The fitted values are: $N_a = 0.00714 \pm 0.00056$, $\mu_{OPD} = 174 \pm 11.4$~nm and $\sigma_{OPD} = 286 \pm 1.01$~nm. Unlike for GLINT data (left), the fitted null depth on this LBTI data set needs a calibration process with a reference star as the instrument simulator does not capture all the noise biases.}
    \label{fig:histos}
\end{figure}

\texttt{GRIP} has successfully measured null depth on already-published data from two very different instruments: GLINT (H-band, spectral dispersion, photonic-based, AO correction, simultaneous measurements of photometries, dark and bright fringes) and LBTI (N band, fringe tracking and AO correction, bulk optics, sequential acquisition of photometries, dark fringe and thermal background).
This library has successfully processed data on Arcturus \citep{martinod2021} taken with GLINT and on $\beta$~Leo \citep{defrere2016} taken with the LBTI.
For both, the same parameters were retrieved: the astrophysical null depth, and the average and standard deviation of a normal distribution modelling the phase fluctuations.
\texttt{GRIP} used the same models found in their respective articles and the same cost functions.
In particular, the null depth fitted with GLINT is the ratio between the fluxes collected in the dark and bright fringes, while the null depth of the LBTI is defined as the ratio between the flux of the nulled output and a reconstructed bright fringe from photometric data.
The results of the fit are shown in Figure~\ref{fig:histos}.
% \texttt{GRIP} also delivers some diagnostic data to visually assess the reliability of the results.
% This diagnostic is the plot of the histograms of the measured and simulated fluxes.
% Indeed, while the fit of the null depth histograms may look fine, important discrepancies can be found in the diagnostic data.
% The Figure~\ref{fig:diag} shows the diagnostic plots for both cases.
% For GLINT, we can see the simulator has some trouble reproducing the histogram of the bright fringe in the 1565~nm.
% For the LBTI data, the histograms of the reconstructed bright fringe and its simulated counterpart are not consistent: the simulation assumes the photometries are constant hence a constant value for the bright fringe.

\begin{figure}[h]
    \centering
    \begin{tabular}{cc}
        \includegraphics[width=0.49\textwidth]{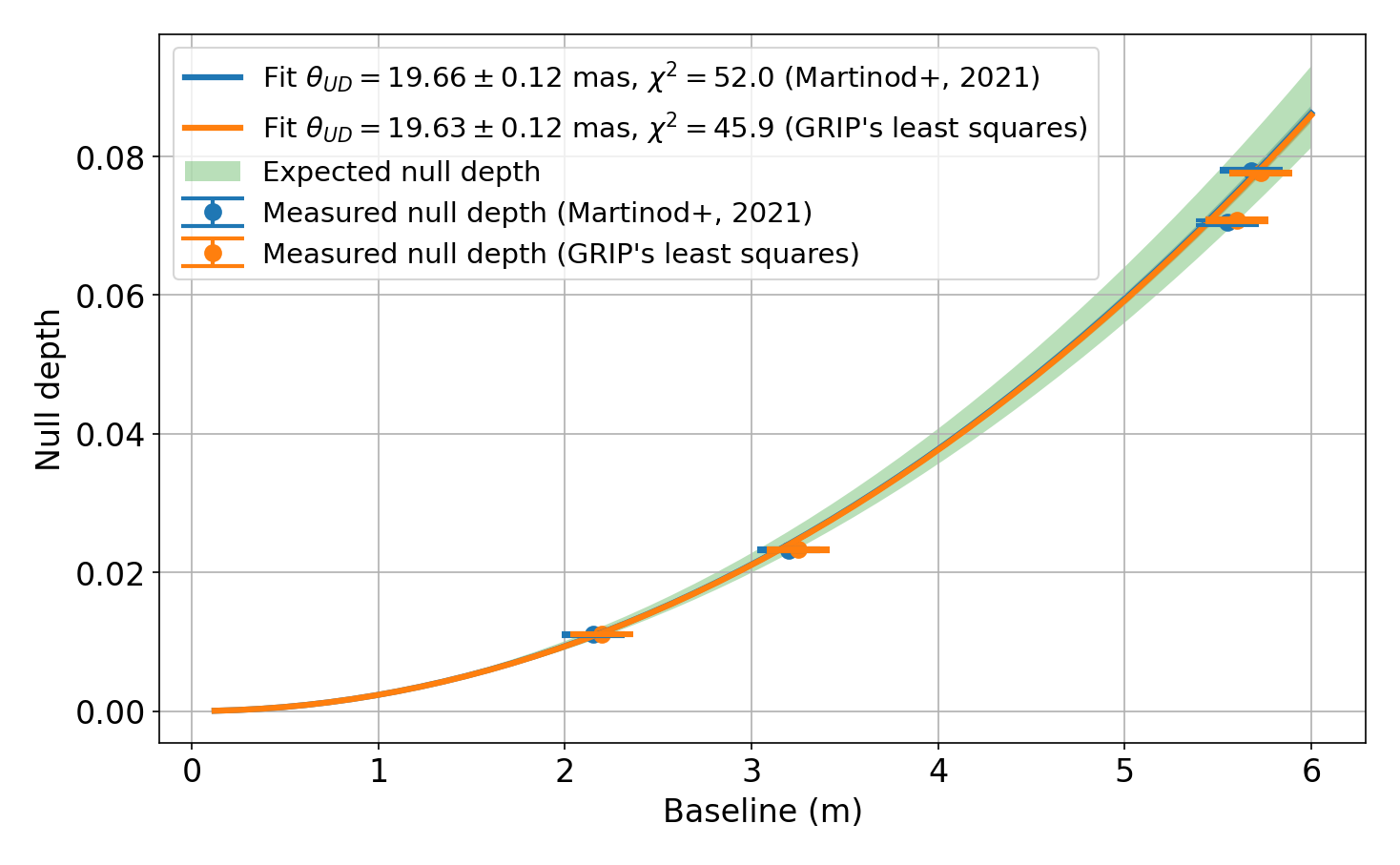} & 
        \includegraphics[width=0.49\textwidth]{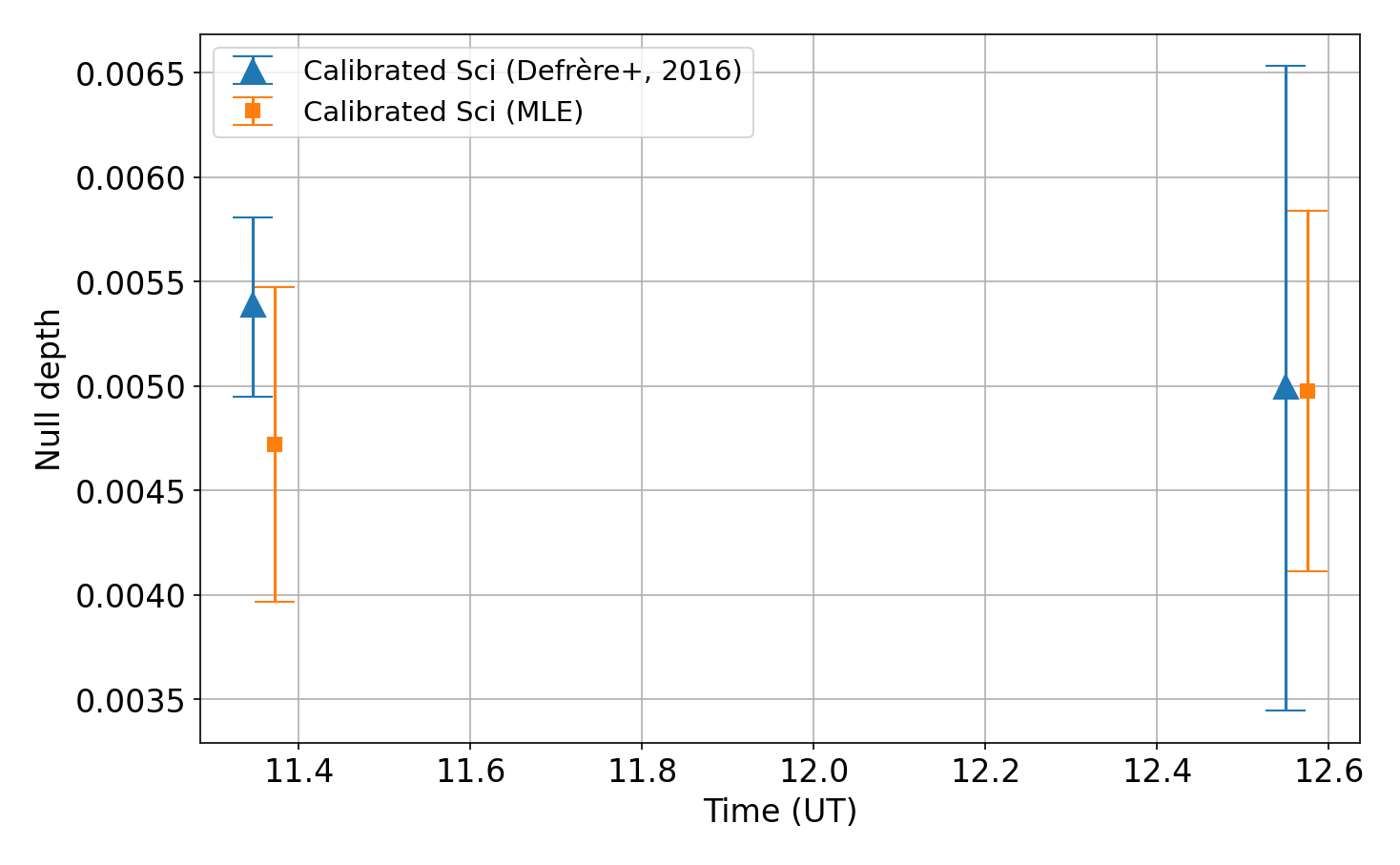} \\
    \end{tabular}
    \caption{Comparison of the results gotten from GRIP with the published data on GLINT (left) and LBTI (right) data. On each figure, the data points from \texttt{GRIP} are artificially horizontally shifted to clarify the plots.}
    \label{fig:final}
\end{figure}

Self-calibrated null depths from \texttt{GRIP} are completely consistent with the published values for GLINT and LBTI (Fig.~\ref{fig:final}).
The null depths given by the package with the least squares method provide the same angular diameter of Arcturus on GLINT data.
Likewise, the null depths on $\beta$~Leo obtained with \texttt{GRIP} with the maximum likelihood estimation are consistent with the published values.

\section{Conclusion and prospects}
\texttt{GRIP} is a modular package providing methods to reduce data from any nuller according to the null self-calibration method.
It has been proven to provide reliable results through several optimisation strategies and to be compatible with any kind of nuller.
This universality is possible by providing a simulator of the instrument along with the data, then using \texttt{GRIP} features to exploit these inputs.
This framework provides the basis for a standardised way of reducing nulling data.
Immediate application will be the NOTT nuller \citep{defrere2018,defere2022spie} on the VLTI in the Asgard suite and GLINT \citep{martinod2021} on the SCExAO bench.
Despite the different architectures of these instruments for performing nulling, \texttt{GRIP} will be able to reduce their data with the same framework.

\texttt{GRIP} is modular and flexible enough to host new data reduction techniques, including those based on neural networks or hierarchical Bayesian modelling.
The increasing number of nullers in activity, motivating the \texttt{GRIP} project, has also raises the need for a standardised data format, similar to OIFITS; it is called NIFITS format.
It aims to encapsulate and exchange calibrated data from any kind of nullers for astrophysical exploitation.

\section*{Acknowledgments}
M-A.M. has received funding from the European Union’s Horizon 2020 research and innovation programme under grant agreement No.\ 101004719.

D.D. acknowledges support from the European Research Council (ERC) under the European Union's Horizon 2020 research and innovation program (grant agreement No.\ CoG - 866070).

S.E. is supported by the National Aeronautics and Space Administration through the Astrophysics Decadal Survey Precursor Science program (Grant No. 80NSSC23K1473).

% References
\bibliography{report} % bibliography data in report.bib
\bibliographystyle{plainnat}

\end{document}